\documentclass[final,1p,times]{elsarticle}

\usepackage{amsfonts}
\usepackage{graphicx}
\usepackage{epstopdf}
\usepackage{epsfig}
\usepackage{amssymb}
\usepackage{setspace}
\usepackage{caption}
\usepackage{amsfonts}
\usepackage{color}
\usepackage{amsmath}
\usepackage{float}
\usepackage{comment}
\newcommand {\be}{\begin{equation}}
	\newcommand {\ee}{\end{equation}}
\newcommand {\bea}{\begin{array}}
	
	\newcommand {\eea}{\end{array}}
\newcommand{\RN}{Reissner-Nordstrom~}
\newcommand{\pt}{\partial}

\journal{Physics Letters B}

\begin{document}

\begin{frontmatter}



\title{Spontaneous pair production near magnetized Reissner-Nordstrom black holes}


\author{Haryanto M. Siahaan}

\address{Jurusan Fisika, Universitas Katolik Parahyangan,\\Jalan Ciumbuleuit 94, Bandung, Jawa Barat, 40141, Indonesia}

\begin{abstract}
We investigate the pair production near a (near) extremal magnetized Reissner-Nordstrom black hole. The pair production is shown to exist in the extremal state, which can be interpreted as the Schwinger effect due to the strong field under consideration. To show a correspondence between the growth of the external magnetic field and the scalar absorption, some numerical examples are provided.

\end{abstract}




\end{frontmatter}


\section{Introduction}\label{sec.intro}
\label{sec:intro}

Vacuum fluctuation near a black hole can yield a pair production of particles in the forms of Hawking radiation and Schwinger effect. It is known that Hawking radiation can be understood in several ways, such as pair production \cite{Hawking:1975vcx} and the tunneling effect \cite{Parikh:1999mf}. However, it was found that Hawking radiation ceases to exist when the black hole reaches extremality. Nevertheless, in such circumstances, pair production can still occur in a way that it resembles a phenomenon known in quantum theory known as the Schwinger effect \cite{Schwinger:1951nm} where a pair of particle and antiparticle is produced from vacuum under an influence of strong field. The similar effect in the vicinity of black hole is also named in the same way \cite{Chen:2012zn} and has been studied in the rotating charged black hole backgrounds \cite{Chen:2016caa,Siahaan:2019ysk,Chen:2020mqs}. Prior to the studies of Schwinger effect near a black hole, the works in \cite{Pioline:2005pf,Kim:2008xv} are addressed to study this effect in the (anti)-de Sitter spacetime.

Recently, astronomical observation suggests the existence of the magnetic field around black hole \cite{EventHorizonTelescope:2021srq}. This observation invites further theoretical studies on black holes that are immersed in external magnetic fields. It has been shown that interaction between black holes and magnetic fields can be performed in two ways. Firstly by considering the external magnetic field as some perturbations as introduced by Wald in \cite{Wald:1974np}. In this approach, the spacetime metric is independent of the external magnetic field parameter, so a null object will not be affected by the presence of magnetic field. Secondly is by utilizing the Ernst magnetization to get a magnetized black hole solution. The magnetized black hole that comes from an Ernst magnetization can be considered as a black hole in the Melvin universe \cite{Ernst:1976mzr}. The spacetime metrics in this case are functions of the external magnetic field parameter, hence the geodesics of null objects depend on the strength of the magnetic field. Recently, aspects of black holes in Melvin universe have been discussed quite extensively \cite{Astorino:2015lca,Siahaan:2021uqo,Siahaan:2021ypk,Siahaan:2021ags,Ghezelbash:2021lcf}. 
The studies conducted in this paper investigate the magnetized spacetime that arises from the Ernst magnetization. However, when using the Wald's magnetic black hole approach, the result regarding pair production with a dependence on the external magnetic field is not obtained. This can be attributed to the fact that the Wald approach is applicable only in the case of weak magnetic fields. 

The work in this Letter is devoted to find out a correspondence between the external magnetic field and the Schwinger effect near a charged black hole. Previously, a study of pair production near a charged black hole was reported in \cite{Chen:2012zn}. Intuitively, the presence of an external magnetic field should amplify the rate of pair production near a black hole, i.e. more particles are produced as the external magnetic field grows. To proceed, first we have to obtain the near horizon geometry of a near extremal magnetized \RN black hole, and then study a massive charged scalar field in that region. The possibility of pair production can be presented by showing the Breitenlohner-Freedman bound violation in the radial equation for the scalar wave. To support the prediction for pair production from the Breitenlohner-Freedman bound violation, one can obtain the Bogoliubov coefficients from the radial solution of the wave equation. It is known that these coefficients correspond to the vacuum persistence and mean number of produced pairs. 

The organization of this Letter is as follows. In the next section, we obtain the near horizon geometry for the near extremal magnetized \RN black hole. The near horizon scalar field is discussed in section \ref{sec.near}, and the associated radial solutions are presented in section \ref{sec.radsol} that also contains some numerical evaluations for the scalar absorption near the black hole. Finally, we give conclusions. In this Letter, we consider the natural units $c={\hbar} = k_B = G_4 = 1$.

\section{Magnetized \RN spacetime and the extremality}

\RN solution solves the Einstein-Maxwell equations
\be \label{eq.EinsteinMaxwell}
R_{\alpha \beta }  = 2F_{\alpha \mu } F_\beta ^\mu   - \frac{1}{2}g_{\alpha \beta } F_{\mu \nu } F^{\mu \nu } \,,
\ee 
where the field strength tensor $F_{\mu\nu} = \partial_\mu A_\nu - \partial_\nu A_\mu$ obeys the source free condition
\be \label{eq.sourcefree}
\nabla _\mu  F^{\mu \nu }  = 0\,,
\ee
and the Bianchi identity
\be \label{eq.Bianchi}
\nabla _{[\mu } F_{\alpha \beta ]}  = 0\,.
\ee\
The line element is given by
\be \label{eq.metricRN}
ds^2  =  - \frac{{\Delta _r }}{{r^2 }}dt^2  + \frac{{r^2 dr^2 }}{{\Delta _r }} + r^2 \left( {\frac{{dx^2 }}{{\Delta _x }} + \Delta _x d\varphi ^2 } \right)\,,
\ee 
whereas the vector field is
\be 
A_\mu dx^\mu = \frac{Q}{r} dt\,.
\ee 
In the equations above, $\Delta_r = r^2 - 2Mr +Q^2$, and $\Delta_x = 1-x^2$. This metric (\ref{eq.metricRN}) describes the spacetime outside a static electrically charged object in Einstein-Maxwell theory with mass $M$ and electric charge $Q$. 

It turns out that the \RN solution can be magnetized by using the Ernst magnetization transformation \cite{Ernst:1976mzr}. The line element and vector solution that describe the magnetized \RN spacetime can be written as
\be \label{eq.metricMagRN}
ds^2  =  - \frac{{\Delta _x \Delta _r }}{f}dt^2  + \frac{{e^{2\gamma } }}{f}\left( {\frac{{dr^2 }}{{\Delta _r }}+\frac{{dx^2 }}{{\Delta _x }}} \right) + f\left( {d\varphi  - \Omega dt} \right)^2 \,,
\ee 
and
\be \label{eq.AmuMagRN}
A_\mu  dx^\mu   = \frac{{Q\left( {1  + f_1 b^2  + f_2 b^4  + f_3 b^6 } \right)dt + br\left( {g_0  + g_1 b^2 } \right)d\varphi }}{{r\left( {1  + 2 g_0 b^2  + g_1 b^4 } \right)}}\,,
\ee 
respectively, where $e^{2\gamma} = r^4 \Delta_x$,
\be
f = \frac{r^2 \Delta_x}{1  + 2 g_0 b^2  + g_1 b^4}\,,
\ee 
\be 
\Omega = \frac{4bQ \left({b}^{2}{r}^{2}{x}^{2}+{b}^{2}{r}^{2}-1+{b}^{2}{Q}^{2}{x}^{2}-2{b}^{2}{x}^{2}Mr\right)}{r}\,,
\ee 
\be 
f_1 = 6Mr{x}^{2}-{r}^{2}{x}^{2}-9{Q}^{2}{x}^{2}-5{r}^{2}\,,
\ee 
\be 
f_2 = \left( 2{Q}^{2}{r}^{2}-9{Q}^{4}-{r}^{4}+12Mr{Q}^{2}-4M{r}^{3} \right) {x}^{4}+ \left( 6{r}^{4}-14{Q}^{2}{r}^{2}+4M{r}^{3}\right) {x}^{2}-5{r}^{4}\,,
\ee 
\be 
f_3 = \left( r^2 \Delta_x +{Q}^{2}{x}^{2} \right) ^{2} \left( x^2 \Delta_r +{r}^{2} \right)\,,
\ee
\be 
g_0 = r^2 \Delta_x +3{Q}^{2}{x}^{2}\,,
\ee 
and
\be 
g_1 = \left( r^2 \Delta_x +{Q}^{2}{x}^{2} \right) ^{2}\,.
\ee
The magnetized \RN solution in the first order of magnetic field parameter is given in eq. (11) of \cite{Dokuchaev:1987ova}. This magnetized solution describes a charged object embedded in the Melvin universe with the external magnetic field parameterized by $b$. The presence of an external magnetic field does not change the extremal radius for the black hole, so the extremal state for the \RN black hole and its magnetized version are just the same at $Q=M$. The Taub-NUT extension of the solution (\ref{eq.metricMagRN}) and (\ref{eq.AmuMagRN}) is presented in \cite{Siahaan:2021ags} where the related Kerr/CFT holography is worked out.

The near horizon of magnetized \RN black hole has been discussed in \cite{Astorino:2015lca} and \cite{Bicak:2015lxa}, related to the Kerr/CFT correspondence and Meissner effect discussions, respectively. In this paper, we employ a near horizon transformation for the near extremal magnetized \RN black hole that generalizes the near horizon transformation employed in \cite{Chen:2012zn} for the non-magnetized \RN geometry. The transformation can be expressed as
\be 
r \to Q + \varepsilon \rho ~~,~~t \to \frac{\tau }{\varepsilon }~~,~~\varphi  \to \phi  - \frac{{4b\left( {1 - b^2 Q^2 } \right)}}{\varepsilon }\tau \,.
\ee 
Together with the coordinate transformation above, we can also consider
\be
M = Q + \frac{{\varepsilon ^2 D^2 }}{{2Q}}\,,
\ee 
where $D$ is a small parameter that measures how far the system is from extremality. After taking $\varepsilon \to 0$, this transformation leads to the near horizon geometry of the near extremal magnetized \RN spacetime that can written as
\be \label{eq.metricnear}
ds^2  = \Gamma \left( x \right)\left( { - \frac{{\left( {\rho ^2  - D^2 } \right)d\tau ^2 }}{{Q^2 }} + \frac{{Q^2 d\rho ^2 }}{{\left( {\rho ^2  - D^2 } \right)}} + \frac{{Q^2 dx^2 }}{{\Delta _x }}} \right) + \frac{{Q^2 \Delta _x }}{{\Gamma \left( x \right)}}\left( {d\phi  - k\rho d\tau } \right)^2 
\ee 
where
\be 
\Gamma \left( x \right) = 1 + b^2 Q^2 \left( {2 + 4x^2  + b^2 Q^2 } \right)\,,
\ee 
and
\be 
k = \frac{{4b}}{Q}\left( {1 + b^2 Q^2 } \right)\,.
\ee 
Note that the near horizon limit requires $\varepsilon\to 0$, whereas the extremality is the state at $D=0$. The spacetime metric (\ref{eq.metricnear}) contains the warped AdS$_3$ that normally appears in the near horizon of near extremal charged and rotating black hole \cite{Chen:2016caa,Siahaan:2019ysk,Guica:2008mu}, instead of taking the AdS$_2\times $S$^2$ in the case of \RN discussed in \cite{Chen:2012zn}. It can be understood from the rotating nature of magnetized \RN spacetime. The corresponding near horizon gauge field in the near extremal geometry can be written as
\be\label{eq.AmuNear}
A_\mu  dx^\mu   = \frac{r}{{Q\Gamma \left( x \right)}}\left( {1 - b^2 Q^2 } \right)\left( {4b^2 Q^2 x^2  - \left( {1 + b^2 Q^2 } \right)^2 } \right)d\tau  - \frac{b}{{\Gamma \left( x \right)}}\left( {Q^2  + b^2 Q^4  + 2Q^2 x^2 } \right)d\phi  \,.
\ee 
The metric solution (\ref{eq.metricnear}) together with the vector (\ref{eq.AmuNear}) solve the equations (\ref{eq.EinsteinMaxwell}), (\ref{eq.sourcefree}), and (\ref{eq.Bianchi}).

\section{Near horizon scalar field}\label{sec.near}

Now let us consider a massive charged test scalar in this near horizon and near extremal state setup. The general form of Klein-Gordon equation for the test scalar field $\Psi$ is
\be \label{eq.KGgen}
\left( {\nabla _\mu   + iqA_\mu  } \right)\left( {\nabla ^\mu   + iqA^\mu  } \right)\Psi  = \mu ^2 \Psi \,.
\ee 
For the scalar field, we employ the separable ansatz $\Psi = e^{-i\omega \tau +i m \phi}R\left(\rho\right)X\left(x\right)$. It yields the Klein-Gordon equation (\ref{eq.KGgen}) to take the form
\[
\frac{1}{R}\partial _\rho  \left[ {\left( {\rho ^2  - D^2 } \right)\partial _\rho  R} \right] + \frac{{Q^2 \left( {Q\omega  + q\rho  + 3q\rho b^2 Q^2  - 4m\rho b\left( {1 + b^2 Q^2 } \right)} \right)}}{{\left( {\rho^2  - D^2 } \right)}} + \frac{1}{X}\partial _x \left[ {\Delta _x \partial _x X} \right]
\]
\be 
- Q^2 \left[ {\left( {1 + b^2 Q^2 } \right)^2  + 4b^2 Q^2 x^2 } \right]\mu ^2  - \frac{{\left( {\Gamma \left( x \right)m - qQ^2 b\left( {1 + b^2 Q^2  + 2x^2 } \right)} \right)^2 }}{{\Delta _x }} = 0\,.
\ee 
The last equation can be separated into two equations that are related by the separation constant $\lambda$, i.e.
\be \label{eq.rad}
\partial _\rho  \left[ {\left( {\rho ^2  - D^2 } \right)\partial _\rho  R} \right] + \left\{ {\frac{{Q^2 \left( {Q\omega  + q\rho  + 3q\rho b^2 Q^2  - 4m\rho b\left( {1 + b^2 Q^2 } \right)} \right)^2}}{{\left( {\rho^2  - D^2 } \right)}} - Q^2 \left( {1 + b^2 Q^2 } \right)^2 \mu ^2 } \right\}R = \lambda R\,,
\ee 
and
\be \label{eq.ang}
\partial _x \left[ {\Delta _x \partial _x X} \right] - \left\{ {\frac{{\left( {\Gamma \left( x \right)m - qQ^2 b\left( {1 + b^2 Q^2  + 2x^2 } \right)} \right)^2 }}{{\Delta _x }} + 4b^2 Q^4 x^2 \mu ^2 } \right\}X =  - \lambda X\,.
\ee

One can consider eq. (\ref{eq.rad}) as an equation for a massive scalar probe propagating in an AdS$_2$ with an effective mass $\mu _{{\rm{eff}}}$ and AdS radius $L_{\rm AdS} = Q$. Following \cite{Chen:2012zn}, the corresponding effective mass can be expressed as
\be \label{eq.effmass}
\mu _{{\rm{eff}}}^{\rm{2}}  = \mu ^2  - \left( {q\left[ {1 + 3b^2 Q^2 } \right] - 4mb\left( {1 + b^2 Q^2 } \right)} \right)^2  + \frac{\lambda }{{Q^2 }}\,.
\ee 
One can observe that this effective mass gets some contributions from the external magnetic field parameterized by $b$. The presence of the quantum magnetic number $m$ in the last expression is understood to give a discrete structure of effective mass in the radial equation (\ref{eq.rad}). It resembles the Zeeman effect for atoms under a magnetic field influence. Indeed, this discreteness that comes from a magnetic field contribution does not appear in the effective mass of a test scalar in the \RN black hole background as discussed in \cite{Chen:2012zn}.

\begin{figure}[h]
	\begin{center}
	\includegraphics[scale=0.25]{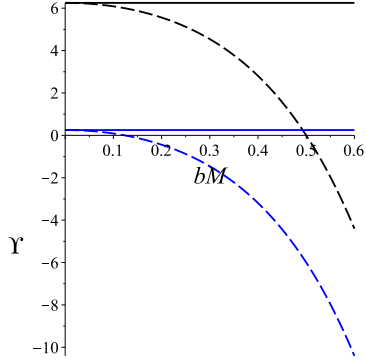}\caption{Illustration for the violation of BF bound for a scalar in the region of magnetized \RN black hole. Here, $\Upsilon = \mu ^2 Q^2  - \left( {q\left[ {1 + 3b^2 Q^2 } \right] - 4mb\left( {1 + b^2 Q^2 } \right)} \right)^2 Q^2   + \left(l+\frac{1}{2}\right)^2$. To get these plots, we consider $Q=M$, $\mu = 10^{-5} M^{-1}$, and $l=2$. The black curves associate to the neutral scalars, whereas the blue ones for charged scalars with $q=10^{-6}$. Solid lines correspond to the case of $m=0$, whereas dashed lines represent $m=1$.}\label{BF}
\end{center}
\end{figure}

It is known that there exist a lower bound for a scalar mass in the $d+1$ dimensional anti-de Sitter spacetime, namely the Breitenlohner-Freedman (BF) bound, that reads
\be 
\mu_{\rm eff}^2 \ge -\frac{d^2}{4L_{\rm AdS}^2}\,.
\ee 
Violation to this bound leads to an instability which can be interpreted to cause a pair production from vacuum. For the effective mass (\ref{eq.effmass}), violation to the BF bound is denoted by
\be \label{eq.BFMag}
\left(\mu ^2  - \left( {q\left[ {1 + 3b^2 Q^2 } \right] - 4mb\left( {1 + b^2 Q^2 } \right)} \right)^2\right) Q^2  + \left(l+\frac{1}{2}\right)^2 < 0\,,
\ee
which associates to the Schwinger pair production and/or Hawking radiation in the vicinity of a near extremal magnetized \RN black hole. Note that the limit $b=0$ of (\ref{eq.BFMag}) gives the condition of particle production in the region close to the near extremal \RN black hole
\be \label{eq.BFRN}
\left(\mu^2 - q^2\right) Q^2 + \left(l +\frac{1}{2}\right)^2 < 0
\ee 
obtained in \cite{Chen:2012zn}. Obviously, the inequality (\ref{eq.BFRN}) can be satisfied for the charged particles only. However, in the magnetized \RN case whose pair production condition is given in (\ref{eq.BFMag}), the neutral particles can be produced as well. 
The production process bears resemblance to that of the Higgs bosons pair production at the LHC \cite{Gouzevitch:2020rwo}. The extreme strength of the external magnetic field, in addition to the gravitational force, prompts the vacuum to produce both charged and neutral particles. Furthermore, this feature can distinguish the pair production of the magnetized \RN black hole and its non-magnetized counterpart. An illustration for the BF bound violation for a scalar in the background of a near extremal magnetized \RN black hole is given in fig. \ref{BF}. It can be seen how the presence of magnetic fields can force the production of neutral scalars, and the correspondence of magnetic field strength to the BF bound violation.

\section{Radial solution and scalar absorption}\label{sec.radsol}

The flux of a probe charged scalar field obeying the Klein-Gordon equation (\ref{eq.KGgen}) can be written as
\be 
{\cal F} = i \int {dxd\phi } \sqrt {\left| g \right|} g^{\rho \rho } \left( {\Psi \tilde \nabla _\rho  \Psi ^*  - \Psi ^* \tilde \nabla _\rho  \Psi } \right)\,,
\ee
where $\tilde \nabla _\mu = \nabla_\mu + i q A_\mu$. Explicitly for the near horizon system discussed previously, this flux takes the form
\be \label{eq.Flux}
{\cal F} = i \Xi \left(r^2 -D^2\right) \left(R \pt_r R^* - R^* \pt_r R\right)
\ee 
where $\Xi = 2\pi \int dx X^* X$.
To discuss the absorption related to the scalar production discussed in the previous section, there exist two kinds of boundary conditions that can be applied. Nevertheless, both boundary conditions are equivalent \cite{Chen:2012zn}. Hence here we will just discuss one of them, namely the outer boundary condition. In this consideration, we set the incoming flux to be vanished at the asymptotic outer boundary. Accordingly, the flux conservation requires
\be 
 \left| {\cal F}_{\rm incident} \right| =  \left| {\cal F}_{\rm reflected} \right| +  \left| {\cal F}_{\rm transmitted} \right|\,,
\ee 
which leads to the Bogoliubov relation
\be 
\left| \alpha \right|^2 = 1 + \left| \beta \right|^2\,,
\ee 
where $\alpha$ and $\beta$ are the Bogoliubov coefficients. The relations between these Bogoliubov coefficients with the fluxes are
\be \label{eq.alpha}
\left| \alpha \right|^2 = \frac{ {\cal F}_{\rm incident}}{ {\cal F}_{\rm reflected}}\,,
\ee 
and
\be \label{eq.beta}
\left| \beta \right|^2 = \frac{ {\cal F}_{\rm transmitted}}{ {\cal F}_{\rm reflected}}\,.
\ee
Here, it is understood that $\left| \alpha \right|^2$ is the vacuum persistence amplitude and $\left| \beta \right|^2$ is the mean number of produced pairs. The absorption cross section then can be written as
\be 
\sigma_{\rm abs} = \frac{ {\cal F}_{\rm transmitted}}{ {\cal F}_{\rm incident}} = \frac{\left|\beta\right|^2}{\left|\alpha\right|^2}\,.
\ee 

To find the corresponding Bogoliubov coefficients related to the particle production near the magnetized \RN horizon, first we have to solve the radial equation (\ref{eq.rad}). The solution to this equation can be written as
\[
R\left( \rho  \right) = C_1 F\left( {\frac{1}{2} + ip + in, \frac{1}{2} - ip + in,1 + i\left( {n + s} \right);\frac{1}{2} - \frac{\rho }{{2D}}} \right)\left( {\rho + D} \right)^{\frac{{i\left( {n - s} \right)}}{2}} \left( {\rho - D} \right)^{\frac{{i\left( {n + s} \right)}}{2}} 
\]
\be 
+ C_2 F\left( {\frac{1}{2} + ip - is,\frac{1}{2} - ip - is,1 - i\left( {n + s} \right);\frac{1}{2} - \frac{\rho }{{2D}}} \right)\left( {\rho + D} \right)^{\frac{{i\left( {n - s} \right)}}{2}} \left( {\rho - D} \right)^{ - \frac{{i\left( {n + s} \right)}}{2}} \,,
\ee 
where
\be 
p = \left\{ {\left[ {\left( {q + 3qb^2 Q^2  - 4mb\left( {1 + b^2 Q^2 } \right)} \right)^2  - \mu ^2 \left( {1 + b^2 Q^2 } \right)} \right]Q^2  - \left( {l + \frac{1}{2}} \right)^2 } \right\}^{1/2} \,,
\ee 
\be 
n = \left( {q + 3qb^2 Q^2  - 4mb\left( {1 + b^2 Q^2 } \right)} \right)Q\,,
\ee
\be 
s = \frac{{\omega Q^2 }}{D}\,,
\ee 
and $F\left(c_1,c_2,c_3;c_4\right)$ is a hypergeometric function. Near horizon, one can approach the radial solution above to become
\[
R\left( \rho  \right) \approx C_1 \left( {\rho + D} \right)^{\frac{{i\left( {n - s} \right)}}{2}} \left( {\rho - D} \right)^{\frac{{i\left( {n + s} \right)}}{2}}  + C_2 \left( {\rho + D} \right)^{\frac{{i\left( {n - s} \right)}}{2}} \left( {\rho - D} \right)^{ - \frac{{i\left( {n + s} \right)}}{2}} 
\]
\be 
= \left( {D} \right)^{\frac{{i\left( {n - s} \right)}}{2}} \left( C_h^{\left( {{\rm{out}}} \right)}  \left( {\rho - D} \right)^{\frac{{i\left( {n + s} \right)}}{2}}  + C_h^{\left( {{\rm{in}}} \right)} \left( {\rho - D} \right)^{ - \frac{{i\left( {n + s} \right)}}{2}} \right) \,,
\ee 
where the identity $F\left(c_1,c_2,c_3;0\right)=1$ has been used. The coefficients $C_h^{\left( {{\rm{out}}} \right)}$ and $C_h^{\left( {{\rm{out}}} \right)}$ are $C_1$ and $C_2$, respectively. At the boundary $\rho \gg D$, one can make use of the related identities in eqs. (\ref{eq.app1}) and (\ref{eq.app2}) to get the approximated radial solution
\be 
R_b \left( \rho \right) \approx C_b^{\left( {\rm in} \right)} \rho^{ - ip  - {\textstyle{1 \over 2}}}  + C_b^{\left( {\rm out} \right)} \rho^{ip  - {\textstyle{1 \over 2}}} \,,
\ee 
with the corresponding constants
\be \label{eq.Cbin}
{C_{ b}^{\left( {\rm in} \right)}} =   {\Gamma \left( { - 2i{p} } \right)}\left[{\frac{{\left( {2D} \right)^{{\textstyle{1 \over 2}} + i\left( {n + {p} } \right)} \Gamma \left( {1 + i\left( {n +  {s} } \right)} \right)C_1}}{{2\Gamma \left( {{\textstyle{1 \over 2}} + i\left( {n - {p} } \right)} \right)\Gamma \left( {{\textstyle{1 \over 2}} + i\left( { {s}  - {p} } \right)} \right)}}} +   {\frac{{\left( {2D} \right)^{{\textstyle{1 \over 2}} - i\left( { {s}  - {p} } \right)} \Gamma \left( {1 - i\left( {n +  {s} } \right)} \right)C_2}}{{2\Gamma \left( {{\textstyle{1 \over 2}} - i\left( {n + {p} } \right)} \right)\Gamma \left( {{\textstyle{1 \over 2}} - i\left( { {s}  + {p} } \right)} \right)}}}\right]\,,
\ee
and
\be \label{eq.Cbout}
{C_{ b} ^{\left( {\rm out} \right)}} =  {\Gamma \left( {2i{p} } \right)} \left[{\frac{{\left( {2D} \right)^{{\textstyle{1 \over 2}} + i\left( {n - {p} } \right)} \Gamma \left( {1 + i\left( {n +  {s} } \right)} \right)C_1}}{{2\Gamma \left( {{\textstyle{1 \over 2}} + i\left( {n + {p} } \right)} \right)\Gamma \left( {{\textstyle{1 \over 2}} + i\left( { {s}  + {p} } \right)} \right)}}}  +  {\frac{{\left( {2D} \right)^{{\textstyle{1 \over 2}} - i\left( { {s}  + {p} } \right)} \Gamma \left( {1 - i\left( {n +  {s} } \right)} \right)C_2}}{{2\Gamma \left( {{\textstyle{1 \over 2}} - i\left( {n - {p} } \right)} \right)\Gamma \left( {{\textstyle{1 \over 2}} - i\left( { {s}  - {p} } \right)} \right)}}} \right]\,.
\ee 

Using the formula (\ref{eq.Flux}), the following fluxes can be computed as
\be
{\cal F}_{h}^{\left({\rm in}\right)} = -2D \Xi \left| C_2 \right|^2 \left(n+s\right)\,,
\ee 
\be
{\cal F}_{h}^{\left({\rm out}\right)} = 2D \Xi \left| C_1 \right|^2 \left(n+s\right)\,,
\ee 
\be
{\cal F}_{b}^{\left({\rm in}\right)} = -2 \Xi \left| C_{b}^{\left({\rm in}\right)} \right|^2\,,
\ee 
and
\be
{\cal F}_{b}^{\left({\rm out}\right)} = 2 \Xi \left| C_{b}^{\left({\rm out}\right)} \right|^2\,.
\ee 
Furthermore, to get the explicit Bogoliubov coefficients, one needs to apply the boundary condition to the fluxes above. As we have mentioned previously, we impose the outer boundary condition here. It can be done by setting ${\cal F}_{b}^{\left({\rm in}\right)}  =0$ that represents the zero incoming flux at the outer boundary. Such consideration yields
\be \label{eq.C1C2}
{C_1} =  - {C_2}{\left( {2D} \right)^{\tfrac{1}{2} - i\left( { s  + n} \right)}}\frac{{\Gamma \left( {1 - i\left( {n +  s } \right)} \right)\Gamma \left( {{\textstyle{1 \over 2}} + i\left( {n - p } \right)} \right)\Gamma \left( {{\textstyle{1 \over 2}} + i\left( { s  - p } \right)} \right)}}{{\Gamma \left( {1 + i\left( {n +  s } \right)} \right)\Gamma \left( {{\textstyle{1 \over 2}} - i\left( {n + p } \right)} \right)\Gamma \left( {{\textstyle{1 \over 2}} - i\left( { s  + p } \right)} \right)}}\,.
\ee 
Moreover, we can have
\be 
{\cal F}_{\rm incident} = {\cal F}_h^{\left({\rm out}\right)}~~,~~{\cal F}_{\rm reflected} = {\cal F}_h^{\left({\rm in}\right)}~~,~~{\cal F}_{\rm transmitted} = {\cal F}_b^{\left({\rm out}\right)}\,,
\ee 
Using the relation in eq. (\ref{eq.C1C2}) in eq. (\ref{eq.Cbout}) can give us
\be 
C_b ^{\left( {out} \right)} =  - {C_2}{\left( {2D} \right)^{\tfrac{1}{2} - i\left( { s  + p } \right)}}\frac{{\Gamma \left( {1 - i\left( {n +  s } \right)} \right)\Gamma \left( {2ip } \right)\sinh \left( {2\pi p } \right)\sinh \left( {\pi \left( {n +  s } \right)} \right)}}{{\Gamma \left( {{\textstyle{1 \over 2}} - i\left( {n - p } \right)} \right)\Gamma \left( {{\textstyle{1 \over 2}} - i\left( { s  - p } \right)} \right)\cosh \left( {\pi \left( {n - p } \right)} \right)\cosh \left( {\pi \left( { s  - p } \right)} \right)}}\,.
\ee 

\begin{figure}[h]
	\begin{center}
		\includegraphics[scale=0.3]{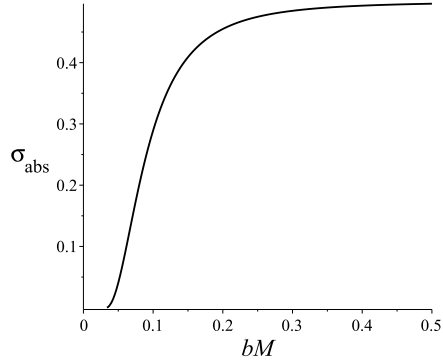}\caption{$\sigma_{\rm abs}$ evaluated for $Q=10$, $\mu = 10^{-2}$, $q=5\times 10^{-3}$, $m=0$, and $l=0$.}\label{sig1}
	\end{center}
\end{figure}

\begin{figure}[h]
	\begin{center}
		\includegraphics[scale=0.3]{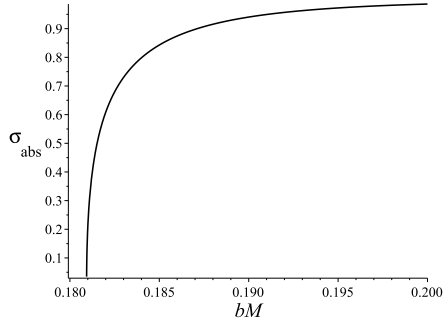}\caption{$\sigma_{\rm abs}$ evaluated for $Q=10$, $\mu = 10^{-2}$, $q=0$, $m=1$, and $l=1$.}\label{sig2}
	\end{center}
\end{figure}

Now, by utilizing the results in (\ref{eq.alpha}) and (\ref{eq.beta}), the vacuum persistence amplitude and the mean number of produced pairs for magnetized \RN black hole can be expressed as
\be \label{eq.alpha.gen}
\left| \alpha  \right|^2  = \frac{{{\cal F} _{\rm incident} }}{{{\cal F} _{\rm reflected} }} = \frac{{{\cal F} _h ^{\left( {out} \right)} }}{{{\cal F} _h^{\left( {in} \right)} }} = \frac{{\cosh \left( {\pi \left( {n + {p} } \right)} \right)\cosh \left( {\pi \left( { {s}  + {p} } \right)} \right)}}{{\cosh \left( {\pi \left( {n - {p} } \right)} \right)\cosh \left( {\pi \left( { {s}  - {p} } \right)} \right)}}\,,
\ee
\be \label{eq.beta.gen}
\left| \beta  \right|^2  = \frac{{{\cal F} _{\rm transmitted} }}{{{\cal F} _{\rm reflected} }} = \frac{{{\cal F} _\infty ^{\left( {out} \right)} }}{{{\cal F} _h^{\left( {in} \right)} }} = \frac{{\sinh \left( {2\pi {p} } \right)\sinh \left( {\pi \left( {n +  s } \right)} \right)}}{{\cosh \left( {\pi \left( {n - {p} } \right)} \right)\cosh \left( {\pi \left( { s  - {p} } \right)} \right)}}\,,
\ee
respectively. Accordingly, the absorption cross section takes the form
\be \label{eq.sigma.gen}
{\sigma _{abs}} = \frac{{\sinh \left( {2\pi {p} } \right)\sinh \left( {\pi \left( {n +  s } \right)} \right)}}{{\cosh \left( {\pi \left( {n + {p} } \right)} \right)\cosh \left( {\pi \left( { s  + {p} } \right)} \right)}}\,.
\ee 
To get the expressions above, we have used the identities in (\ref{eq.app3}). Indeed, the results in (\ref{eq.alpha.gen}), (\ref{eq.beta.gen}), and (\ref{eq.sigma.gen}) above still get a contribution from Hawking radiation. Nevertheless, in an extremal state, the pair production can only occur in the form of Schwinger effect. This state is achieved by taking $D\to 0$ which yields $s\to \infty$. Accordingly, the vacuum persistence amplitude and mean number of produced particles can be written as
\be\label{eq.alphaExt}
\left| \alpha  \right|^2 = \frac{{\cosh \left( {\pi \left( {n + p } \right)} \right)}}{{\cosh \left( {\pi \left( {n - p } \right)} \right)}}\exp \left( {2\pi p } \right)~~,~~
\left| \beta  \right|^2  = \frac{{\sinh \left( {2\pi p } \right)}}{{\cosh \left( {\pi \left( {n - p } \right)} \right)}}\exp \left( \pi\left({n + p }\right) \right) \,,
\ee
which correspond to the absorption cross section
\be \label{eq.sigmaExt}
\sigma _{abs}  = \frac{{\sinh \left( {2\pi p } \right)}}{{\cosh \left( {\pi \left( {n + p } \right)} \right)}}\exp \left( \pi\left({n - p }\right) \right)\,.
\ee 
The results above agree to that appearing in \cite{Chen:2012zn} after setting $b\to 0$ and $q \to -q$. This scalar absorption cross section belongs to the Schwinger effect in the vicinity of an extremal magnetized \RN black hole. To illustrate the contribution of the external magnetic field to the effect, we evaluate the scalar absorption cross section in (\ref{eq.sigmaExt}) for some numerical setups in figs. \ref{sig1} and \ref{sig2}. From fig. \ref{sig1}, we can learn that the growth of the magnetic field yields an increase of $\sigma_{\rm abs}$. On the other hand, fig. \ref{sig2} tells us that the neutral test particles can be produced as a Schwinger effect near the magnetized black hole, for the state of non-vanishing quantum numbers $m$ and $l$. Likewise, for the production of neutral scalar, $\sigma_{\rm abs}$ increases as the magnetic field becomes larger.

\section{Conclusion}\label{sec.conc}

We have shown that an external magnetic field around an extremal charged black hole can amplify the Schwinger effect in the vicinity of its horizon. This result is in line with our expectation, where the rate of particle production should increase as the magnetic field gets stronger \cite{Schwinger:1951nm}. Interestingly, for the case of magnetized black holes, the pair production can occur for both charged and neutral particles. Previously, it was reported that the pair production near \RN black hole \cite{Chen:2012zn} can exist only for the charged particles. 

For future work, we would like to investigate the significance of other physical parameters in the Plebanski-Demianski family of black hole solutions to the rate of pair production near the horizon. Note that considering the rotational parameter $a$ for the pair production calculation has been performed in \cite{Chen:2016caa}. In addition to the electric charge and rotation, the black hole spacetime solutions in the Plebanski-Demianski family can consist of acceleration and NUT parameters as well \cite{Griffiths:2009dfa}. Finding out the correlation between these additional parameters to the Schwinger effect should be interesting.

\appendix

\section{Some useful identities}

A transformation formula for the hypergeometric function can be written as
\be \label{eq.app1}
 F\left( {a,b;c; \frac{D-\rho}{{2D}}} \right) = \frac{1}{2} \left[ \left( {\frac{{\rho + D}}{{2D}}} \right)^{ - a} F\left( {a,c - b;c;\frac{{\rho - D}}{{\rho + D}}} \right) + \left( {\frac{{\rho + D}}{{2D}}} \right)^{ - b} F\left( {b,c - a;c;\frac{{\rho - D}}{{\rho + D}}} \right)\right]\,.
\ee 
The Gauss's summation theorem can be expressed as
\be \label{eq.app2}
F\left(a,b;c;1\right) = \frac{\Gamma\left(c\right)\Gamma\left(c-a-b\right)}{\Gamma\left(c-a\right)\Gamma\left(c-b\right)}\,,
\ee 
for $\Re\left(c\right) > \Re\left(a+b\right)$. For the Gamma function, we have
\be \label{eq.app3}
\left| {\Gamma \left( {\frac{1}{2} + ix} \right)} \right|^2  = \frac{\pi }{{\cosh \pi x}}~~,~~\left| {\Gamma \left( {1 + ix} \right)} \right|^2  = \frac{\pi }{{\sinh \pi x}}~~,~~\left| {\Gamma \left( {ix} \right)} \right|^2  = \frac{\pi }{{x\sinh \pi x}}\,.
\ee

\end{document}